\documentclass[12pt]{article}

 \usepackage{graphicx}

\begin{document}

\centerline{\large\bf Acoustic Edge Magnetoplasmons and }

\centerline{\large\bf  Quantum Hall Effect}

 \centerline{ V.B.Shikin}
\centerline{Institute of Solid State Physics, Russian Academy of
Sciences,}

\centerline{142432 Chernogolovka, Russia}

\vskip 3 mm \
\begin{abstract}

Discussed in the present study are details of the behavior of low-frequency ``acoustic'' ($ac$) modes in the spectrum
of edge magnetoplasma oscillations in axially symmetric degenerate 2D electron systems where electron density
distribution $n(\vec r)$ behaves as $n(\vec r \to R)\to 0$ when $r$ approaches the external radius $R$ of the domain
occupied by electrons. It is shown that finding the dependence of the spectrum of $ac$-modes on radial ($l$) and
azimuthal ($m$) indices when both $l$ and $m$ are small requires axially symmetric solution of the relevant problem.
The desires is achieved in the so-called elliptic approximation for electron density distribution $n(\vec r)$. The
obtained results are employed to interpret available data on the excitation of $ac$-modes in degenerate electron disks
with smooth electron density profile placed in the magnetic field $H$ normal to the disk plane. The performed analysis
confirms reported detection of the soft $ac$-mode in the range of $H \gg H_{max}$ where $H_{max}$ is the field at which
the maximum of the $\omega_{lm}^{s}(H)$ curve is observed (here and below $\omega_{lm}^{s}(H)$ stands for the frequency
of the soft $ac$-mode).

A strong interaction of the $ac$-modes with $(integer)$-channels inevitably arising near the boundaries of 2D electron
systems with smooth electron density profile as the magnetic field is varied in the Quantum Hall Effect (QHE) regime is
emphasized. Well-formed $integer$-stripes can suppress some acoustic modes which is actually observed in experiments.

\end{abstract}

PACS:

\baselineskip=20pt

\centerline{\textbf{Introduction}}

The existence of special low-frequency modes (later called ``acoustic'') in the spectrum of edge magnetoplasma (edge
magnetoplasmons, EMP) oscillations was predicted in Refs. [1, 2]. Generally, $EMP$-oscillations propagate along the
boundary of a 2D charged system placed in the magnetic field $H$ normal to its plane. The existence of ``acoustic''
($ac$; $AEMP$) modes requires vanishing of the charge density $n(\vec r)$ at the disk boundary. Special attention to
the condition  $n(r)_{| r=R} \to 0 $ (here $R$ is the 2D disk radius) in the studies of $EMP$ is easily understood. A
necessary component in the formulation of all classical problems involving edge plasma oscillation is the requirement
of vanishing normal component of induced current at the sample boundary
$$
    j_n(r\to R) = 0 .\eqno(1)
$$
The charge density $n(\vec r)$ in Eq. (1) is assumed to be finite up to the boundary $\vec r=R $. However, if the
boundary is ``soft'', i.e.
     $$
         n(r)_{| r=R} \to 0 \eqno(2)
     $$
(for example, in 2D systems with externally controlled structure), Eq. (1) is satisfied automatically which should have
a profound effect on the spectrum in such systems.

Calculations of the details of classical $EMP$ spectrum in the system with ``elliptic'' density profile (the term
``elliptic'' is explained below, see Ref. [2] and Eq. (7)) confirms this guess. For a disk with finite radius $R$, in
addition to the discrete plasmon spectrum modified by the cyclotron splitting due to magnetic filed $H$, a new acoustic
mode arises whose frequency raises from zero at zero $H$ (hence the term ``acoustic'') to some peak value at a certain
$H$ where
$$
\partial \omega_{ac}(H=H_{max})/\partial H =0, \eqno(3)
 $$
after which $\omega_{ac}$ decreases approximately as $1/H$. $AEMP$-excitations were discovered [3,4] in a disk of
surface ions in liquid helium exactly at the stage of their frequency going through the maximum as a function of
magnetic field $H$ which substantially simplified their identification. Various attempts (listed in the recent
published studies [5--7] on the $\omega_{ac}$ problem) to perform similar experiments in degenerate electron systems
have not yet been successful.

In practice, the expected for the $\omega_{ac}$ frequency range reveals a set of modes whose frequency diminishes
monotonously with growing magnetic field which is typical of all varieties of $EMP$. Actually, an essential argument in
favor of the $ac$-nature of the excitations observed in Ref. [7] is the inequality
$$
\frac{\omega_{ac}(q, j)}{\omega_{ac}(q, j+1)} > 1, \eqno(4)
$$
(here $q$ is the wave number along the disk boundary, $j$ is the discrete index from the definition of $\omega_{ac}(q,
j)$ (5)) which should be satisfied by $ac$-modes with different indices. Usually, the discrete $EMP$-excitations have
the energy which increases with the growth of the azimuthal and radial indices. $AEMP$ modes behave in quite the
opposite way as indicated by Eq. (4).

Inequality (4) is present in all versions [1, 2, 4, 8] of the
$ac$-modes description. The solution [8] for the half-plane with
the boundary where density profile $n(x)$ becomes zero: $n(x)_{|x
\to 0} \to 0$ reveals this point in the most prominent way. A
special choice of the density profile $n(x)$ in the half-plane
problem allows to obtain an analytic expression for the spectrum
of $ac$-modes
$$
\omega_j(q)= -s_j q, \quad s_j = \frac{2\bar n_s e^2}{\epsilon m_e
\omega_c j}, \quad  j =1, 2, 3 ... \eqno(5)
$$
Here $\epsilon$ is the effective dielectric constant, $m_e$ is the effective mass, $\omega_c$ is the cyclotron
frequency, $\bar n_s$ is the average density of the 2D system far from the transition domain, $q$ is the wave number
along the ``soft'' boundary. In the representation (5) $\omega_j(q)$ reveals no peaks monotonously decreasing with the
magnetic field $H$, thus indicating that Eq. (5) is only applicable for sufficiently high fields exceeding the field at
which the ``acoustic'' peak occurs. However, the dependence $s_j \propto j^{-1}$ which is typical of the $ac$-modes
and, consequently, validity of the observed inequality (4) is present in the results of Ref. [8].

Observation of the property (4) derived from Eq. (5) is assumed to be sufficient proof of the ``acoustic'' nature of
experimentally detected excitations. However, the results of Ref. [9] cast a doubt on this confidence. Here the
$Hall-bar$-geometry  and $time ~resolved$ techniques were employed to detect the excitations whose velocities also
comply with Eq. (5). The authors of Ref. [9] believe their measurements to be relevant to predictions of Ref. [8].
However, in the cell geometry used in Ref. [9] the requirement (2), which is a necessary condition for the existence of
$ac$-modes (for details, see below), cannot be satisfied. In this connection, additional arguments [7] (mainly
experimental) allowing to assume the existence of a smooth density profile possessing special properties (2) become
important.

The point is that as the magnetic field approaches the range of $H
\ge H_{max}$, the classical regime (5) changes to the threshold
suppression of $ac$-modes in the fields approximately
corresponding to integer filling factors $\nu_l= integer $ ~ ($\nu
=\pi l_{H_l}^2 \bar n_s, ~~ l_H^2=2c\hbar/eH $ ) within the
homogeneous part of the 2D system at the disk center [5-7]. This
the way the acoustic modes $\omega_{ac}^{j}(H)$ respond to the QHE
state of the 2D system. In the present paper general arguments of
Ref. [7] concerning possible reasons of strong influence of the
QHE on $ac$-modes are filled with specific contents allowing to
understand the details of observed transformation of $ac$-modes.
This part of the study is based on modification of the classical
formalism [1, 2, 4, 8] employing the simplest possible approach to
description of inhomogeneous systems, namely, local Drude
approximation. The limits of this model by today's standards was
discussed in detail in Refs. [10, 11] for Ohmic transport
$\sigma_{ik}^{\parallel}(x)$ in the direction normal to the
electron density gradient. Modification of the conduction
properties $\sigma_{ik}^{\perp}(x)$ of inhomogeneous 2D system
along the direction of $d n/dx$ proves to be equally important.
Details of this modification are directly related to the
properties of $ac$-modes in the QHE regime, as shown in the
present paper.

$$
$$

\centerline{\textbf{1.}  Classical $EMP$-excitations in the disk with elliptic density profile}

\textbf{A.} In order to present a review of available results we first consider the main criteria (4), (5) of
``acoustic'' nature of the modes observed in Ref. [7] in axially symmetric terms. Any ``flat'' statement for edge
excitations (including inequality (4)) is only meaningful for cylindrical geometry in the limit
$$
    \lambda \ll R,\eqno(6)
$$
where $\lambda$ is the edge excitation wavelength, $R$ is the
characteristic radius of the studied disk. The relation given by
Eq. (4) combined with Eq. (5) does not depend on $q$ at all,
suggesting either its universal nature with respect to requirement
(6) or (which is more probable) its quantitative unsuitability in
axially symmetric problems. The only possibility to find out the
real state of things arises if an appropriate axially symmetric
solution is available. In the present case this possibility is
provided [2] by the known solution for the $EMP$-excitations in
the disk with the elliptic  density profile defined as
$$
    n(r)= n(0)\sqrt{(1-r^2/R^2)}, \eqno(7)
$$
which is a suitable alternative to the profiles  $n(x)$  employed in Refs. [1, 8].

The classical spectrum $\omega(l, m, \omega_c)$ of $EMP$-excitations in the disk with the density profile (7) and
dissipationless tensor $\sigma_{ik}(r, \omega)=\sigma_{ik}(\omega) n(r)/n(0)$ within the Drude approximation
$$
\sigma_{xx}(\omega)=\frac{i\omega n(0)
e^2}{m_e(\omega^2-\omega_c^2)}, \quad
\sigma_{xy}(\omega)=\frac{\omega_c n(0)
e^2}{m_e(\omega^2-\omega_c^2)}, \quad \omega_c= \frac{eH}{m_e
c}\eqno(8)
$$
($n(r)$ from (7), $m_e$ is the free electron mass, $c$ is the speed of light) normalized according to Ref. [2]
$$
\Omega_{lm} = \Omega_0/L_{lm}^{1/2},\quad  L_{lm} = 2
\frac{\Gamma(l + m + 1)}{\Gamma(l + m + \frac{1}{2})}
               \frac{\Gamma(l + 1)}{\Gamma(l + \frac{1}{2})},\quad  \Omega_0^2 =
               \frac{2\pi n(0) e^2}{\varepsilon m_{{\rm e} } R},\eqno(9)
 $$
($\varepsilon$ is the ambient media dielectric constant, $\Gamma(x)$ is the gamma-function) has the following
structure:
$$
     \omega_{lm}^2 - [\omega_c^2 + (2l + m)(2l + m + 1) - m^2] = m(\omega_c/\omega_{lm}).  \eqno(10)
$$
Here  $l$, $m$ are the radial and azimuthal indices taking any integer values starting from zero. To stress the
difference between spectra (5) and (10), radial index in (10) is denoted by symbol $l$.

For $m > 0$ (the case of $m = 0$ means zero wave numbers and therefore is irrelevant) the follwing two parts of the
dispersion equation (10) are of interest: the versions with $l = 0$ and $l > 0$. For $l = 0$ one of the roots of Eq.
(10) coincides (to within its sign) with the cyclotron frequency $\omega_{0m}= -\omega_c$. This root should be ignored
since when deriving Eq. (10) the original dispersion equation was multiplied by $(\omega^2 - \omega_c^2)$ [2]). the two
remaining roots yield the frequencies (in usual units, s$^{-1}$)
$$
    \omega^{\pm}_{0m} = \sqrt{\frac{|m|}{L_{0m}}\Omega_0^2+\frac{\omega_c^2}{4} }\pm\frac{\omega_c}{2} , \quad
     L_{0m} =   \frac{2}{\sqrt{\pi}} \frac{\Gamma(m + 1)}{\Gamma(m + \frac{1}{2})}, \eqno(11)
   $$
with $\Omega_0^2$  from (9).

In the problem of $ac$ - excitations the properties of clearly
observed $EMP$ modes (11) (shown in Fig. 1) are useful due to
their adjusting possibilities employed below.

For $l > 0$, Eq. (10) is a cubic equation with respect to $\omega_{lm}$ and has three real roots correspnding to
magnetoplasma oscillations of two types. First, there exist two modes whose frequencies $\omega_{lm}^{\pm}$ remain
finite if $\omega_c \to 0$
$$
    \omega_{lm}^{\pm}(\omega_c \to 0) \to
    \Omega_0\sqrt{A_{lm}}/L_{lm}^{1/2}, \quad A_{lm} = [(m + 2l)(m +
    2l + 1) - m^2].\eqno(12)
$$
In the limit $\omega_c \to \infty$ ~ $\omega_{lm}^{\pm}$ asymptotically approach $\omega_c$ from above.

Second, there is a low-frequency $ac$-mode ($AEMP$) with the frequency first increasing with the growth of $H$ as
$$
    \omega_{lm}^{s} \simeq m\omega_c/A_{lm}, \eqno(13)
$$
and then, passing through a peak (3), decreasing as $\omega_c^{-1}$
$$
    \omega_{lm}^{s} \simeq \frac{\Omega_0^2}{\omega_c}\frac{m}{L_{lm}}\eqno(14)
$$
at $\omega_c \gg 1 $ (dimensionless units (9)).

In the range of $l \gg 1$ the constants $L_{lm}$ in (9) can approximately can be calculated as
$$
    L_{lm} \simeq  2 \sqrt{l(l + m)},\eqno(15)
$$
and therefore the limit of large $l$ at fixed $m$ in Eq. (14)
yields
$$
    \omega_{lm}^{s} \simeq \frac{\pi n(0) e^2 m}{\varepsilon m_e \omega_cR\sqrt{l(l+m)}}.\eqno(16)
$$
Hence, for $l \gg m$ the asymptotic relations (5) and (16) both have the same structure: $j \leftrightarrow l$, and the
relevant parameters are practically identical. Everything indicates a qualitative agreement between (5) and (16) in the
limit (6).

As to the general case, Eq. (14) yields an axially symmetric alternative for the ``flat'' inequality (4)
$$
\omega_{l,m}^{s}/\omega_{l,(m+1)}^{s} =
\frac{mL_{l,m+1}}{(m+1)L_{l,m}} < 1,\quad
\omega_{l,m}^{s}/\omega_{(l+1),m}^{s} =
\frac{L_{l+1,m}}{L_{l,m}}>1 \eqno(17)
$$
where $L_{l,m}$ is from (9).

\textbf{B.} Papers [5-7] do not contain description of the details of the mechanism used to excite different modes.
Hence the only way to identify the indices actually governing the spectra is to analyze the behavior of the spectra
themselves. One of stages of this analysis concerns the choice of index $m$. Let us consider the available data.
Analysis of the upper right inset in Fig. 1 [7] reveals that in the $j$-representation the ratio of the frequencies
$\omega_{j=1}^{exp}/\omega_{j=2}^{exp} \le 1.9 $ (approximately 1.8--2.0, the accuracy of this numeric interval is
rather low) for not too high magnetic fields (where no signs of suppression are yet observed) and does not depend on
$m$ (i.e., on $q$) at all. In the $l$-representation given by Eq. (14) this observable ratio leads, according to (17),
to the requirement (18) expressed as an equation for $m$
$$
\frac{L_{lm}(l=2, m)}{L_{lm}(l=1, m)}\simeq 1.9.\eqno(18)
$$

Trying in Eq. (18) various possible small values of index $m$, one
arrives at the following result: the case of $m=0$ is irrelevant,
$m=1$ yields $L_{lm}(l=2, 1)/L_{lm}(l=1, 1)= 1.6$. The rest values
$m>1$ only reduce this ratio which tends to unit for large $m$.
The index $m=1$ provides the value closest to that given by Eq.
(18). Not too impressive numerical agreement between 1.9 and 1.6
can be explained by non-ellipticity of real disks used in Refs.
[5--7].

\begin{figure}[tbp]%%%%%%%%%%%1
\begin{center}
\includegraphics*[width=6 cm]{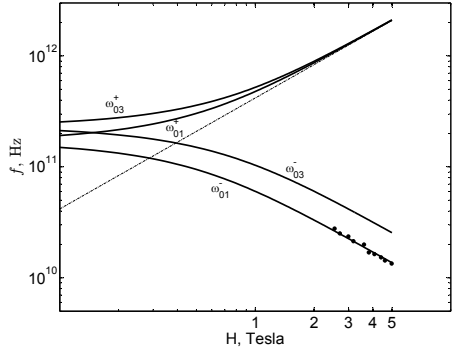}
\end{center}
\caption{ Magnetoplasmon frequencies (10), (11) with $m = 1, ~ m = 3 $, $l = 0$ for the disk with elliptic equilibrium
electron density profile. Dashed line: $\omega = \omega_c$. The choice of $m = 1$ is described in the comments to Eq.
(18), while index $m = 3$ just demonstrates the increase in the energy of the $EMP$-mode with growing indices (an
alterative to Eq. (4), the frequency $\omega_{0m}^{-}$ increases with $m $). The rest numbers are given in the text in
the comments to the figure. Shown in the lower branch are the experimental data for $ \omega^{-}_{01}(H)$ from [5].
Adjustment was performed with the parameter $n(0)$ as described in the text.}
\end{figure}
\begin{figure}[tbp]%%%%%%%%%%%1
\begin{center}
\includegraphics*[width=6 cm]{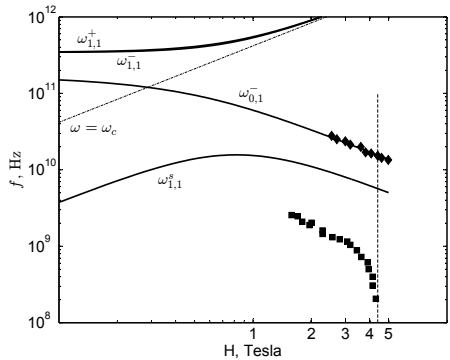}
\end{center}
\caption{Magnetoplasmon spectrum $\omega_{lm}$ (10),(12) for $ m = 1, ~l = 1$. Two upper curves corresponding to
$\omega_{1,1}^{\pm}$ have a slight linear in $H$ splitting at $H \to 0 $ and the lower of them has a shallow minimum
practically indiscernible in the figure. The lowest curve representing the $AEMP$-mode $\omega_{11}^{s}$ with the
numbers identical to those of $EMP$-plots in Fig. 1 illustrates position of the $AEMP$-peak with respect to the range
of magnetic fields studied in Ref. [7]. The experimental data for $AEMP$-modes [7] are shown by squares. Plotted for
comparison is the curve $\omega^{-}_{01}(H)$ from Fig. 1 with experimental points (diamonds) used to normalize all the
numbers in Figs. 1 and 2. Vertical dash-and-dot line indicates the threshold above which the $AEMP$-mode vanishes. The
same line indicates that the $EMP$ mode $ \omega^{-}_{01}(H)$ is practically insensitive to the threshold.}
\end{figure}

Frequencies $f=\omega/(2\pi)$ in units of s$^{-1}$ of elliptic $EMP$-modes (10), (11) with indices $(l = 0, m = 1)$,
$(l = 0, m = 3)$ and normalization (9) are presented in Fig. 1. The plots are drawn taking into account data of Figs. 1
and 2 from Ref. [5]: disk radius $R = 15$ $\mu$m, electron mass $m_e = 0.067m_0$, effective dielectric constant of the
ambient medium $\varepsilon = 6.9$.  The field $B$ is replaced with $H$ since $\mu \simeq 1$, electron density at the
disk center $n(0) = 6.3\times 10^{11}$ cm$^{-2}$ (used as adjustable parameter to fit the experimental frequencies [5]
ignoring the density of $2.2 \times 10^{11} $ cm$^{-2}$ corresponding to the average sample density before etching).
The relevant experimental points lie on the curve $\omega_{01}^{-}$ in the lower right part of the figure. The curve
$(l = 0, m = 3) $ is plotted to demonstrate an increase in the energy of $EMP$-modes with growing indices (an
alternative to Eq. (4)).

$AEMP$-part of the problem is represented numerically by plots in Fig. 2 where the indices $m=1$ , $l = 1$, are used
combined with the normalization (9); the required numerical disk parameters are the same as those adopted for
calculations of $EMP$-modes plotted in Fig. 1. The upper part of Fig. 2 contains information on frequencies
$f=\omega_{lm}^{\pm}/(2\pi)$ following from (10) and covering asymptotic behavior given by Eq. (12). The lower branch
$f=\omega_{lm}^{s}/(2\pi)$ is drawn according to Eq. (10) for the mode with $l=1, m = 1 $ within the range of magnetic
fields where the peak for $ac$-excitations can coexist on a single figure with the experimental points (black squares)
reported in Fig. 1of Ref. [7].

In addition to the $AEMP$-part of the spectrum, Fig. 2 contains $EMP$-resonances $  \omega^{-}_{01}(H)$ (black
diamonds) also shown in Fig. 1. The purpose of this juxtaposition is to demonstrate the absence of visible signs of
suppression of $EMP$ modes at magnetic fields where the $AEMP$ excitations already no longer exist. The observed
selectivity (some modes vanish while others persist) is characteristic of the mechanisms of the mode suppression
discussed in section \textbf{2.}

A noticeable difference in Fig. 2 between the observed and calculated frequencies $f=\omega_{11}^{s}/(2\pi)$ obtained
by fitting of the $EMP$-resonances $ \omega^{-}_{01}(H)$ is somewhat disappointing. On the other hand, it is
appropriate to remind the original reasons of turning to the axially symmetric formalism in the description of the
$AEMP$ excitation against the background of predictions [8] for the semi-infinite model of the 2D system. There were
some arguments suggesting a qualitative difference in the structure of inequalities (4) and (17) that were actually
confirmed later. A possibility arose for a thorough analysis of the azimuthal indices $m$ whose values are actually
determined by the details of the $EMP$-modes excitation technique. In the final results of Ref. [5] this index is a
free parameter. A justification was developed for dealing with the frequencies themselves rather only their ratios. As
a result, a qualitatively understandable picture presented in Fig. 2 was derived where the position of branches
$\omega_{11}^{s}$, $ \omega^{-}_{01}$ relative to the line $\omega=\omega_c$ is shown and the peak location $H=H_{max}$
(3) is indicated. One can hardly hope for the elliptic model to claim for anything more. As to the indicated numerical
discrepancy, it originates not only from the poor agreement of the density profile adopted in the elliptic model (7)
and the real disks. The data of Ref. [5] for the $EMP$-part of spectrum reveal that changing the nominal disk size 4
times (from 5 to 20 mcm) only reduces the frequency of the $\omega^{-}_{01}$-mode by a factor of 2 which is quite
unexplainable within the traditional theory of $EMP$-excitations [20]. As a consequence this puzzle is also transferred
to the $AEMP$-domain.

One more reason for numerical disagreement in frequencies is the deviation of the $\omega_{lm}^{s}(H)$ curve from the
asymptotic law (14) in the entire range of experimentally studied magnetic fields [7], as shown in Fig. 2. In this
domain the $ac$-modes start feeling their ```end'' of quantum origin marked in Fig. 2 by a vertical line. Quantitative
description of the $\omega_{lm}^{s}(H)$ behavior in the vicinity of the threshold is still missing. The origin of the
threshold itself is discussed below in Section \textbf{2.}

$$
$$

\centerline{ \textbf{2.} Influence of integer stripes on the dynamics of $ac$-modes}

\textbf{A.}  Typical structures developing on the density profile $n(r)$ in the QHE regime are flat areas with $d n/dx
=0$ (integer channels) in the vicinity of points $x_l$ where $\nu(x_l)=integer$
$$
 \nu(x)= \pi l_{H_l}^2 n(x), , \quad l_H^2=2c\hbar/eH\eqno(19)
$$
Chklovsky et al. [12, 13]). Fig. 3 depicts two such stripes located at the points where $\nu(x_2) =2$ and $\nu(x_4) =4$
that were calculated self-consistently in Refs. [14--16]. The left axis measures the local filling factor $\nu(x)$ from
(19) for the electron density profile $n(x)$ as a function of position $x$ (solid line in the figure) and local value
of $\sigma_{xx}^{\perp}(x)$ (dash-and-dot line) at the same points. The density profile $n(x)$ of a one-dimensional
symmetrical [$n(x)=n(-x)$] about the origin $x=0$ two-dimensional system considered in Ref. [16] has clearly defined
zeros at the system boundaries [$n(b)=n(-b)=0$]. Position of movable zero points $\pm b$ relative to the sandwich
boundaries $\pm d$ is controlled by external parameters described in detail in Refs. [14--16]. Fig. 3 is drawn for
$d=1.5 \mu m $, $b=0.9 d$. When magnetic field normal to the plane of 2D system is gradually raised it finally reaches
a threshold value $H_{thresh}$ sufficient for development of as single (and for higher fields even multiple) integer
shelf on the density profile. Shown in Fig. 3 are two such shelves. The first one corresponds to $\nu_l =2$, while the
second has $\nu_l =4$. Even values of the filling factor are due to the particular approximation adopted in Ref. [16]
(neglect of the electron spin).
\begin{figure}[tbp]%%%%%%%%%%%1
\begin{center}
\includegraphics*[width=4 cm]{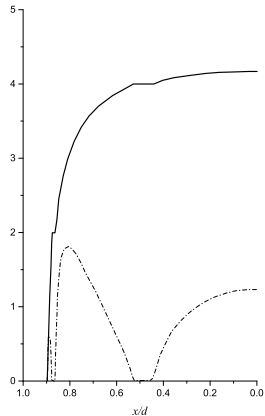}
\end{center}
\caption{Filling factor $\nu(x)$ (solid line) containing flattened
areas where $d\nu(x)/dx =0$ (integer shelves, the origin of the
term is explained below) and the profiles of the conductivity
$\sigma_{xx}^{\perp}$ in the direction normal to the integer
shelves axes (dash-and-dot line) for the left half of a symmetric
sample with $d= 1.5 \mu m $, $n_s=4\cdot 10^{11} cm^{-2}$ , $b=0.9
d$ calculated within the $TFA-SCBA$ approximation (Thomas-Fermi
approximation employing the self-consistent Born approximation) by
authors of Ref. [16] (a part of Fig. 5 from [16]).}
\end{figure}

The data of Fig. 3 allow to distinguish (as to $EMP$) between smooth and sharp transition domains in the electron
density behavior. The profile $n(x)$ is considered to be smooth if the integer shelf width $2a_l$ at its slope is small
compered with the length scale of the transition domain of $n(x)$. Bearing in mind the definition of $a_l$ from [12]
$$
    a_l^2= 2 \kappa \hbar \omega_c/(\pi^2e^2 dn(x_l)/dx)\eqno(20)
$$
($\kappa$ is the ambient medium dielectric constant, $\omega_c$ is the cyclotron frequency) and the estimate for the
transition domain width $w$ as given by (21)
$$
    w \simeq\int_0^{\infty}dx[n_{\infty}-n(x)]/n_{\infty},
\eqno(21)
$$
one obtains the requirement for smooth profile in the form
$$
    a_l \ll w.\eqno(22)
$$
It should be noted that the very definition of $a_l$ (20) has some limitations. The calculations performed assumed a
possibility for arising electrostatic fields to vanish far from the central part of the shelf located at $x_l$. This
simplification requires careful treatment of the shelf properties near the edges of 2D system [17], [18] (for Fig. 3,
this is the shelf with $\nu =2$).

Apart from confirming the hypothesis of the authors of Refs. [12, 13] concerning the possibility of formation of
$integer$ shelves, the papers of Gerhardts and coauthors [16] proposed a technique for calculation of the conductivity
$\sigma_{xx}^{\perp}(x)$ under inhomogeneous conditions of Fig. 3. According to their calculations (dash-and-dot line
in Fig. 3) the conductivity $\sigma_{xx}^{\perp}(x)$ becomes zero in the areas where $d n(x)/dx =0$. This is most
clearly seen for $\sigma_{xx}^{\perp}((x)$ within the shelf with $\nu_l =4$.

Taking into account the importance of anomalous behavior of the conductivity $\sigma_{xx}^{\perp}(x)$ following from
the calculations [16]
$$
    \sigma_{xx}^{\perp} \equiv \sigma_{xx}(\nu_l)\equiv 0, \eqno(23)
$$
(the conductivity of an $integer$-stripe in the direction normal to its axis is zero), it is useful to mention the
direct experiments [19] favoring Eq. (23).  Reported in the review [19] are data on the IVC of a single
$integer$-stripe in the direction normal to its axis. It is shown that in the Ohmic range it is strongly non-linear in
the sense that
$$
    \partial J/\partial V_{V \to 0} \to 0, \eqno(24)
$$
($J$ is the total current crossing the stripe, $V$ is the potential difference at its sides), which is equivalent to
the property $\sigma_{xx}^{\perp} \simeq 0$ (23).

\textbf{B.} According to Fig. 3, apart from the domain $n(r \to R) \to 0$ which is a necessary component of the
classical theory of $ac$-modes [1, 2, 8] (based on the local Drude definition of the conductivity  $\sigma_{xx} \propto
n(x)$ covering the area where $n(r \to R) \to 0$), the QHE regime features additional domains with prominent modulation
of the 2D system conductivity. In that case classical approximation $\sigma_{xx} \propto n(x)$ fails even
qualitatively.

Quantitative analysis of the QHE-induced features in the behavior of $ac$-excitations due to development of
$integer$-stripes on the density profile $n(x)$ can be performed in two ways: by directly including information of Fig.
3 for $\sigma_{xx}^{\perp}(x)$ into equations describing $EMP$ dynamics; or (without claiming any numbers)
approximately, by replacing the details in the behavior of $\sigma_{xx}^{\perp}(x)$ shown in Fig. 3 with the boundary
conditions prohibiting charge transfer through the shelves.

Let the area $x=x_4$ be the sole $integer$-channel on the density profile $n(x)$ shown in Fig. 3. Then to the right of
point $x_4$ (chosen as the origin) equilibrium electron density is almost uniform so that the set of equations for the
oscillating part of $\delta n_s^>(x,t)$ has the standard structure typical of the $EMP$ formalism [20]. The required
solution $\varphi_>(x,y,t)$ should decay towards the sample center and satisfy the requirement
$$
    j_{\perp}(x \to + 0)=0.\eqno(25)
$$
In the vicinity of $x=0$ electron density is not zero. Hence, condition (25) has the sense of requirement (1). The
difference between (1) and (25) is that the potential $\varphi_>(x,y,t)$ from (1) to the right of point $x=0$ is the
only one relevant to the problem. Its involvement in the boundary condition (1) finally leads to the $EMP$ dispersion
law. The problem (25), in addition to the right part of the 2D system, there exists the left part between the point
$x=x_4$ and the boundary of the system where electron density vanishes. The requirement (25) prevents the charge
exchange between the left and right parts of the $integer$-stripe. However, the mutual influence of the fields
$\varphi_>(x,y,t)$ and $\varphi_<(x,y,t)$ on the conducting areas to the left and to the right of the cut (see below)
contributing to the spectrum formation remains finite.

The potential $\varphi_>(x,y,t)$ for the problem (25) (with the origin shifted to the center of the $integer$-stripe)
is written as

$$
    \varphi_>(x,y,t)=\varphi_>(x)\exp{(iqy)}\exp{(-i\omega
    t)}\eqno(26)
$$
$$
\varphi_>(x) \simeq
    \frac{2\sigma_{xx}}{i\omega\kappa}\int_0^{+\infty}
    K_0(|x-s|)\left[\frac{\partial^2 \varphi_>}{\partial
    s^2}+\frac{\partial^2 \varphi_<}{\partial s^2}\right]ds, \quad
    x\ge 0. \eqno(27)
$$
$K_0(x)$ being the zero order Bessel function of imaginary argument.

To the right of point $x = 0$ the conductivity $\sigma_{xx}$ in (27) has some small efective value different from its
quantum value $\sigma_{xx}=0$ at an integer shelf. Zero $\sigma_{xx}$ in the vicinity of $x= 0$ is accounted for by
introducing a discontinuity in the potential described by two separate functions $\varphi_>(x,y,t)$ and
$\varphi_<(x,y,t)$ appropriately matched at the line $x=x_4\pm 0$.

To the left of the point $x=x_4-0$ (or  $x=0$ for (27)) the 2D electron stripe is squeezed from both sides by zero
normal current conditions. This allows one to consider it as effectively quasi-one-dimensional system responding only
to the fields depending on the $y$ coordinate,
$$
    -i\omega \delta n_l(y) + \sigma_{yy}(\omega,
    H)\left[\frac{\partial^2 \varphi_>}{\partial y^2}+\frac{\partial^2
    \varphi_<}{\partial y^2}\right] \simeq 0, \quad \delta
    n_l(y,t)\simeq \int\limits_0^{x_4}\delta n_s^<(x,y,t)dx \eqno(28)
$$
$$
    \varphi_<(x,y) \simeq e\delta n_l(y) K_0[q(x - a_4)]\eqno(29)
$$
The potential $\varphi_<(x,y)$   (29) is typical of the quasi-one-dimensional conductors with a logarithmic singularity
at the filament axis which is in any case cut off at the length equal to $a_4$ with $a_l$ taken from (20). As a rough
estimate one obtains, assuming in (28) $ \varphi_> \sim 0$ and $\sigma_{yy} \propto (i\omega)^{-1}$, that equations
(28), (29) contain the dispersion law $\omega \propto q$ typical of one-dimensional conductors. The simplification
(28), (29) is not critical. The problem with discontinuity at the line $x=x_4-0$ can also be accurately formulated for
the case of two boundary conditions of different types. On of them has the same sense as (25). The other condition,
classical, preserves the structure of Eq. (2).

The two equations (27), (28) determine the spectrum of $EMP$-mode along the integer stripe located at $x_4$ close to
the edge of the 2D disk with elliptic profile (7). The outer part of this structure is actually a quasi-one-dimensional
ring which partly screens the fields of the $EMP$-mode localized at the inner side of the integer stripe with the
coordinate $x_4$ in Fig. 3.

To solve this equation set, one can try a perturbation theory since $ \varphi_> \propto \sigma_{xy}$ while the
perturbation $\varphi_<(x,y)\propto \sigma_{yy}$ (which verified directly) and additionally the inequality
$\sigma_{yy}/\sigma_{xy} \ll 1$ is assumed to be satisfied. In that case the zero approximation to solution of equation
set (27), (28) satisfies the integro-differential equations
$$
    \varphi_>(x) \simeq
    \frac{2\sigma_{xx}}{i\omega\kappa}\int\limits_0^{+\infty}
    K_0(|x-s|)\frac{\partial^2 \varphi_>}{\partial s^2}ds, \quad x\ge
    0, \quad  d^2\varphi_>/dx^2 \gg d^2\varphi_>/dy^2 \eqno(30)
$$
for $\varphi_>(x)$ with the boundary condition
$$
    \sigma_{xx}\varphi_>^{\prime}(0)+i q
    \sigma_{xy}\varphi_>(0)=0\eqno(31)
$$

Integrating Eq. (30) by parts and then adding and subtracting in the r.h.s of the definition of $\varphi_>(x)$ the
combinations
$$
    K_0(qx)\varphi_>^{\prime}(x)\Longleftrightarrow
    \int\limits_0^{+\infty} \frac{\partial K_0(q|x-s|)}{\partial
    s}\varphi_>^{\prime}(x) ds,
$$
one obtains
$$
    \varphi_>(x) \simeq \frac{2\sigma_{xx}}{i\omega\kappa}\left\lbrace
    K_0(qx)[\varphi_>^{\prime}(0)-\varphi_>^{\prime}(x)]+\int\limits_0^{+\infty}
    \frac{\partial K_0(q|x-s|)}{\partial s}
    [\varphi_>^{\prime}(x)-\varphi_>^{\prime}(s)]ds,\right\rbrace.\eqno(32)
$$

Substitution into the expression (32) for $\varphi_>(x)$ of the value $x=0$ yields
$$
    \varphi_>(0)=\frac{2\sigma_{xx}}{i \omega
    \kappa}\varphi_>^{\prime}(0)\int
    \limits_0^{+\infty}\left[1-\frac{\varphi_>^{\prime}(s)}{\varphi_>^{\prime}(0)}\right]
    \frac{\partial K_0(qs)}{\partial s}ds \eqno(33)
$$

Further analysis of Eq. (33) is based on the hypothesis that
$$
    \frac{\varphi_>(0)}{\varphi_>^{\prime}(0)} \simeq -l , \quad
    \mbox{and} \quad
    \left[1-\frac{\varphi_>^{\prime}(s)}{\varphi_>^{\prime}(0)}\right]\simeq
    \left\{\begin{array}{rcl}0, ~~x <l \\ 1~~x >l.\\\end{array}\right.
    \eqno(34)
$$

Then equations (34) and (33) yield
$$
    -l \simeq \frac{2\sigma_{xx}}{i\omega\kappa} K_0(ql). \eqno(35)
$$
As a consequence, boundary condition (31) together with (34) and (35) result in
$$
    \omega(q) \simeq \frac{2\sigma_{xy}}{\kappa}q K_0(ql), \quad
    \omega \ll \omega_c. \eqno(36)
$$
It is important that the conductivity $\sigma_{xx}$ in the definition (35) for $l$ has a finite value shown by dashed
lines in the vicinity of the edges of the integer stripe in Fig. 3. The zero value of $\sigma_{xx}(\nu_l)\equiv 0$
itself within the integer stripe is accounted for by the boundary conditions (25), (31).

The dispersion law (36) reveals the existence (just as in Ref. [20]) of a standard $EMP$-mode along the
$integer$-stripe whose axis is positioned at $x=x_4$ on the density profile $n(x)$ in Fig. 3. This stament is
consistent with the observed data presented in Fig. 2. Vertical line which marks the threshold for suppression of the
low-frequency $AEMP$-mode $\omega_{11}^{s}$ corresponds to the magnetic field where the $EMP$-mode $\omega_{01}^{-}$
demonstrates quite regular behavior.

\centerline{\textbf{4.} Summary}

Proposed is an interpretation of experimental data [5-7] revealing the existence of $AEMP$-excitations in 2D charged
disks with a ``soft'' profile $n(r)$ of degenerate electron density. The discussion covers asymptotic behavior of such
excitations in strong magnetic fields $H \gg H_{max}$ where $ H_{max}$ is the field at which a characteristic peak in
the $\omega_{ac}(H)$ curve occurs. The analysis is based on the elliptic approximation (7) for electron density profile
$n(r)$. The axially symmetric elliptic approximation is shown to much more realistic in calculations of the observed
$AEMP$-excitations in 2D disks [5-7] than the usually employed ``semi-infinite'' approach. Parameter $\delta$ (21),
which qualitatively measures the ellipticity degree of the problem is, according to data of Ref. [5], $\sim 1$ (for the
semi-infinite geometry ($\subset$-formalism) it should satisfy the inequality $\delta \ll 1$), the observed indices $l$
and $m$ of $AEMP$-excitations in Fig. 2 are close to the minimal possible ones, which also creates difficulties in
applying the half-plane formalism [8] to the observed $AEMP$-dynamics with axial symmetry.

An additional argument favoring $AEMP$-origin of the observed modes is a beautiful effect of their suppression in
magnetic fields $H > H_{thresh}$ which are high enough for development in the vicinity of the 2D disk boundary of
$integer$-stripes typical of QHE. Formally, this results in adding to the equations describing the $EMP$-excitations
one more requirement (25), (31) affecting the $EMP$-dynamics governed by the boundary condition (2). Analysis of
equations (27),(28) accounting for the boundary conditions (2), (31) reveals that in this formulation the $AEMP$-modes
do not survive which is confirmed by experiment.

The author is grateful to S.Nazin for very useful discussions, constructive remarks and help in drawing figures. My
acknowledgements are also to I.Andreev, V.Muraviev, and I.Kukushkin for discussion of general situation in
$AEMP$-problem and useful remarks.

\centerline{References}

1. S.Nazin, V.Shikin, ZhETF,  94, (1988), 133 (in Russian)

2. S.Nazin, V.Shikin, FNT 15, (1989), 227 (in Russian)

3. E.Elliott, C.Pakes, L.Skrbek and W.Vinen, Phys.Rev.Lett, 75,
(1995), 3713

4. E.Elliott, S.Nazin, C.Pakes, L.Skrbek , W.Vinen, G.Cox,
Phys.Rev. B 56, (1997), 3447

5. D.Smetnev, V.Muraviev, I.Andreev, I.Kukushkin,  Pis'ma ZhETF, 94,(2011), 141 (in Russian)

6. I.Andreev, V.Muravev, I.Kukushkin, Pis'ma ZhETF, 96, (2012), 588 (in Russian)

7. I.Andreev, V.Muravev, D.Smetnev, I.Kukushkin,  Phys.Rev B 86, (2012), 125315

8. L.Aleiner, L.Glazman  Phys.Rev.Lett, 72, (1994), 2935

9. G.Ernst, R.Haug,, J.Kuhl, K. von Klitzing, K.Eberl,
Phys.Rev.Lett 77, (1996), 4245

10. V.Shikin, S.Nazin,  ZhETF 113, (2011), 306 (in Russian)

11. S.Nazin and V.Shikin, Quantum Hall effect in narrow Coulomb channels. Physical Review B, 84 (2011) 153301

12. D.Chklovskii, B.Shklovskii, L.Glazman,  Phys. Rev. B 46
(1992), 4026.

13. D.Chklovski, K.Matveev, B.Shklovskii,  Phys. Rev. B 47,
(1993), 12605

14. K.Lier, R.Gerhardts,  Phys.Rev. B 50, (1994), 7757

15. J.Oh, R.Gerhardts,  Phys.Rev. B 56, (1997) 13519

16. A.Siddiki, R.Gerhardts,  Phys.Rev. B 70, (2004), 195335

17. V.Shikin, Yu.Shikina, FTT 39, (1997), 742 (in Russian)

18. V.Shikin, Pis'ma ZhETF, 73, (2001), 283 ; 73, (2001), 605 (in Russian)

19. E.Devyatov, UFN, 177, (2007), 207 (in Russian)

20. A.Mikhailov, in $Horison ~in~ World~ Physics$, edited by O.Kirichek, v.236,  Chap.1 (Nova, New York, 2000), p 1-47

\end{document}